\documentstyle[12pt]{article}
 
\begin{document}


 \begin{center}
{\Large \bf
Chiral-Odd Structure Function $h_1^D(x)$\\
and Tensor Charge of the Deuteron.}\\[1cm]

{\large \bf
 A.Yu.  Umnikov$^{a,b,}$\footnote{INFN Postdoctoral Fellow},
 Han-xin He$^{b,c}$ and F.C. Khanna$^b$}\\[5mm]

{\em $^a$Department of Physics, University of Perugia, and
INFN, Sezione di Perugia,\\
via A. Pascoli, Perugia, I-06100, Italy.}

{\em $^b$University of Alberta, Edmonton, Alberta T6G 2J1, Canada
and\\ TRIUMF, 4004 Wesbrook Mall, Vancouver, BC, Canada, V6T 2A3.}

{\em $^c$China Institute of Atomic Energy, P.O.Box 275(18), Beijing
102413, P.R. China.}
\end{center}

\begin{abstract}
The   chiral-odd
structure function $h_{1}^D(x)$
and the tensor charge of the deuteron
are studied within  the Bethe-Salpeter
formalism for the deuteron amplitude.
Utilizing a simple model for the nucleon structure
function, $h_1^N$,  $h_1^D(x)$
is calculated and the nuclear effects are analyzed.
\end{abstract}

\section{Introduction}

There is an increasing interest in the transverse quark distributions
in the nucleon, especially the chiral-odd structure function of  twist
two~\cite{rs,jj},
$h_1^N(x)$, and its first moment,
the tensor charge of the nucleon~\cite{hj,hj1,js,kpg,it,adhk}.
The reasons for the interest are the following.
First, $h_1^N(x)$, complimentary to the conventional
structure functions $F_{1,2}^N$ and $g_{1,2}^N$,
 is essential to understand the nucleon spin substructure.
At the same time, the tensor charge, $g_T^N$, like
the axial charge, is one of the fundamental observables for the nucleon.
So far, however, little is known about both $h_1^N(x)$ and $g_T^N$.
Second,  semi-inclusive
deep-inelastic experiments aiming to measure $h_1^N$, and hence the
tensor charge of the proton and neutron, $g_T^N$, are
being planned~\cite{exp}.
Now is the time for the theory to make predictions.

 It is important to remember that experimental results for
the neutron can only be obtained from the indirect experiments, using the
the lightest nuclear targets, the deuteron and $^3He$. Therefore,
it is necessary to have a reliable  estimate of the nuclear effects,
in order to extract  dependable information about the neutron
from the nuclear data.

In this letter we study the chiral-odd structure function and the
 tensor charge
of the deuteron, $h_1^D(x)$ and $g_T^D$, in the Bethe-Salpeter formalism.
The main motivation is to study the nuclear effects in the
transverse-polarized deuteron structure function, $h_1^D$, and
 the deuteron tensor charge, $g_T^D$.
Since the nucleon tensor charge measures the net number of transversely
polarized valence quarks (quarks minus antiquarks) in a transversely
polarized nucleon~\cite{hj1}, and so it gets no contribution from the
sea-quarks,
the neutron isoscalar tensor charge is equal to the one of the proton.
As a result, the measurement of the deuteron tensor charge may, in principle,
provide a test for  the models of the nuclear effects if the proton
 tensor charge is known precisely. In addition, an independent measurement
of the neutron's $h_1^n$ and $g_T^n$ is obviously useful to understand the
transverse spin distribution in the nucleon. Therefore it is essential
to account for the nuclear effects in the deuteron
for both the chiral-odd structure function
$h_1^D$ and the tensor charge, $g_T^D$.

\section{Basic formulae}

The chiral-odd structure function of the deuteron of twist two,
 $h_1^D(x)$,
is defined as a simple extension of
the case of spin-half hadrons~\cite{jj}.
It can be expressed
in terms of a matrix element of the quark bilocal operator:
\begin{eqnarray}
\int \frac{d\lambda}{2\pi} e^{i\lambda x} \langle P_D \, S_D \mid
\bar \psi(0) i\sigma^{\mu\nu} \gamma_5 \psi(\lambda n)
 \mid P_D \, S_D \rangle
= 2 h_1^D(x)\cdot T^{\mu\nu}+ ({\rm higher \quad twist\quad terms}).
\label{defh}
 \end{eqnarray}
An antisymmetric tensor, $T^{\mu\nu}$, in the rest frame of the
deuteron  is defined by:
\begin{eqnarray}
T^{\mu\nu}\equiv \left (
S^{\mu}_D P_D^{\nu}- S^{\nu}_D P_D^{\mu}
\right ) ,
\label{deft}
\end{eqnarray}
where $P_D^\mu$ is the deuteron momentum, $S_D^\mu$ is the
 spin of the deuteron~\cite{hjm}:
 \begin{eqnarray}
S_{D}^\mu(M) = -\frac{i}{M_D}\epsilon^{\mu\alpha\beta\gamma}
E^*_\alpha(M)E_\beta(M)P_{D\gamma}, \label{spin1}
\end{eqnarray}
where $E(M)$ is the deuteron polarization, with $M = \pm 1,0$, $M_D$ is
the deuteron mass. $S_D$ satisfies the conditions, $S_D^2 = -1$ and $P_DS_D =
0$.

The tensor charge of the deuteron is the first moment of $h_1^D(x)$
and can be readily obtained
from eq.~(\ref{defh}):
\begin{eqnarray}
  \langle P_D \, S_D \mid
\bar \psi(0) i\sigma^{\mu\nu} \gamma_5 \psi(0)
 \mid P_D \, S_D \rangle
= 2 g_T^D(x) T^{\mu\nu}.
\label{defg}
\end{eqnarray}

Using the following identities in the rest frame of the deuteron:
\begin{eqnarray}
&&\left. T^{\mu\nu}T_{\mu\nu}\right|_{M=1} = -2M_D^2
\label{id1}\\
&&\left. i T^{\mu\nu}\sigma_{\mu\nu}\gamma_5\right|_{M=1} =
 -2M_D \gamma_5\gamma_3\gamma_0,
\label{id2}
\end{eqnarray}
the structure function  $h_1^D(x)$
 and the charge $g_T^D$
are rewritten as the matrix elements of the Lorentz scalar  operators:
\begin{eqnarray}
h_1^D(x) &=& \frac{1}{M_D}\int \frac{d\lambda}{2\pi} e^{i\lambda x} \langle P_D
\, S_D \mid
\bar \psi(0)  \gamma_5\gamma_3\gamma_0 \psi(\lambda n)
 \mid P_D \, S_D \rangle,
\label{defhs}\\
 g_T^D &=& \frac{1}{M_D} \langle P_D \, S_D \mid
\bar \psi(0) \gamma_5\gamma_3\gamma_0 \psi(0)
 \mid P_D \, S_D \rangle,
\label{defgs}
\end{eqnarray}

It is impossible to solve eqs.~(\ref{defhs}) and (\ref{defgs})
directly, since the hadron state, $\mid P_D \, S_D \rangle$,
 still cannot be described in QCD. Instead, we use approach where
the ``elementary'' structure function, $h_1^N$, is obtained on the
quark level, while the nuclear effects are calculated on the hadron level
(e.g.~\cite{uk,mg1} and references therein).
We utilize the Bethe-Salpeter formalism, which was previously applied for
studying the structure functions $F_{1,2}^D$, $b_{1,2}^D$ and
$g_{1}^D$~\cite{uk,ukk}. In what follows, we neglect
small possible off-mass-shell
effects in the nucleon structure functions~\cite{mst,uoff}.

In terms of the Bethe-Salpeter amplitudes $h_1^D$ reads:
\begin{eqnarray}
 &&h_1^D(x) = i
  \int \frac{d^4p}{(2\pi)^4}
{h_1^N}\left( \frac{x m}{p_{10}+p_{13}}\right)
\frac{\left. {\sf Tr}\left\{
\bar\Psi_M(p_0,{\bf p})\gamma_5\gamma_3\gamma_0
 \Psi_M(p_0,{\bf p}) (\hat p_2-m)
\right \}\right |_{M=1}}{2(p_{10}+p_{13})},
 \label{h1}
\end{eqnarray}
where m is the nucleon mass, $p$ is the relative momentum
of  nucleons in the deuteron,
 $p_{10}$ and $p_{13}$ are the time and
3-rd components of the nucleon momentum, $p_1$,
and $p_{1,2}=P_D/2 \pm p$. $\Psi_M(p_0,{\bf p})$ is the Bethe-Salpeter
amplitude for the deuteron with the spin projection $M$
and the isoscalar structure function $h_1^N$ is defined as
$h_1^N (x)\equiv (h_1^p(x) + h_1^n(x))/2$. Note,
that there is no explicit mesonic (sea quarks)
contribution to the structure function $h_1^D$. The mesons manifest
themselves by binding the nucleons in the deuteron and, therefore,
defining the structure of the amplitude $\Psi_M(p_0,{\bf p})$.

Then eq.~(\ref{h1}) can be rewritten in the convolution form~\cite{jaffe}:
\begin{eqnarray}
&& h_1^D(x) = \int\limits_{x}^{M_D/m}
\frac{dy}{y}
h_1^N \left( \frac{x}{y}\right) H^{N/D}(y),
 \label{conh1}
\end{eqnarray}
where $H(y)^{N/D}$ is
the
``effective transverse distribution'' of nucleons in the deuteron
and its definition is obvious from eq.~(\ref{h1}).
This is a usual result for the nuclear structure functions calculated
in the leading twist approximation, neglecting the off-mass-shell
deformation of the nucleon structure function. Eqs.~(\ref{h1})
and (\ref{conh1}) allow immediately to write the sum rule for the
deuteron tensor charge:
\begin{eqnarray}
g^D_T &=& \int\limits_{0}^{1}dx h_1^N(x) \cdot \int\limits_{0}^{M_D/m}
{dy}
 H^{N/D}(y)
 \label{sr1}\\
&=& g_T^N \cdot \langle P_D \, S_D \mid
\bar N(0) \gamma_5\gamma_3\gamma_0 N(0)
 \mid P_D \, S_D \rangle,
\label{sr2}
\end{eqnarray}
where $N(x)$ is the nucleon field.
Eqs.~(\ref{sr1}) and (\ref{sr2}) define the renormalization factor
of the nucleon tensor charge by the deuteron structure.
This sum rule is similar to the sum rule for the deuteron
structure function, $g_1^N$~\cite{ukk}:
\begin{eqnarray}
\int\limits_{0}^{1}dx_D g_1^D(x_D) =
 \int\limits_{0}^{1}dx g_1^N(x) \cdot \langle P_D \, S_D \mid
\bar N(0) \gamma_5\gamma_3 N(0)
 \mid P_D \, S_D \rangle.
\label{sr3}
\end{eqnarray}
This sum rule is used for estimating the integral of the
neutron structure
function, $g_1^n$, from the combined proton and deuteron data.

Note that in the static limit the matrix elements on the r.h.s.
of eqs.~(\ref{sr2}) and (\ref{sr3}) are equal
 and
approximate
formulae for the deuteron structure functions are:
\begin{eqnarray}
 g_1^D(x_D) \simeq \left (1-\frac{3}{2}w_D \right)
g_1^N(x),
\label{sr4}\\
 h_1^D(x_D) \simeq
\left (1-\frac{3}{2}w_D \right) h_1^N(x),
\label{sr5}
\end{eqnarray}
where $w_D$ is weight of the d-wave component in the deuteron
wave function.

\section{Numerical calculations}

The method to calculate numerically   expressions like (\ref{h1})
is discussed in ref.~\cite{ukk}. The most important details of
the calculations are:
\begin{enumerate}
\item A realistic model for the Bethe-Salpeter amplitudes
is essential for a realistic estimate of the
nuclear effects. We use a recent numerical solution~\cite{uk}
of the ladder Bethe-Salpeter equation with a realistic
exchange kernel~\cite{tjond}.
\item The Bethe-Salpeter amplitudes and, therefore, eq.~(\ref{h1})
have a nontrivial singular structure. These singularities
must be  carefully taken into account~\cite{fs,land,km}.
\item The BS amplitudes are numerically calculated with the
help of the Wick rotation. Therefore, the
procedure of the numerical inverse Wick rotation must be applied.
\item An essential ingredient of the calculations,
the nucleon structure function $h_1^N(x)$, is unknown both theoretically
and experimentally.
A reasonable estimate for using in
our calculations should be  found.
\end{enumerate}

The effects of the nuclear structure, the Fermi motion and
binding of nucleons, is encoded in the effective distribution function
$H^{N/D}(y)$ in eq.~(\ref{conh1}).  The result of
calculation of $H^{N/D}(y)$ with the Bethe-Salpeter amplitude
of the deuteron~\cite{uk,ukk} is shown in Fig.~{\ref{H}} (solid line).
Two other effective distributions for the deuteron are
also shown in Fig.~\ref{H} for comparison. The first
(dashed line) is $\vec f^{N/D}(y)$, which defines the spin-dependent
structure function $g_1^D$. Explicit expression for $\vec f^{N/D}(y)$
is similar to the one for $H_1^{N/D}(y)$ (see eqs.~(\ref{h1})
and (\ref{conh1})), but with the operator $\gamma_5(\gamma_0+\gamma_3)$
instead of $\gamma_5\gamma_3\gamma_0$. If we
again replace the operator by $(\gamma_0+\gamma_3)$, we get another
distribution, $f^{N/D}(y)$ (dotted line), which defines the
spin-independent structure function $F_2^D$.
Each of the three effective distribution
functions ($H_1^{N/D}(y)$, $\vec f^{N/D}(y)$
or $f^{N/D}(y)$) would represent
the correspondent structure functions of the deuteron
 ($h_1^D$, $g_1^D$ or $F_2^D$) if the nucleons in the deuteron
would be the  elementary fermions.

Note that the function $H_1^{N/D}(y)$  is smaller at
the maximum than $\vec f^{N/D}(y)$, and also slightly
smaller in its normalisation:
\begin{eqnarray}
&&\int\limits_0^{M_D/m} dyH^{N/D}(y)= \langle P_D \, S_D \mid
\bar N(0) \gamma_5\gamma_3\gamma_0 N(0)
 \mid P_D \, S_D \rangle = 0.9208,
\label{sr2a} \\
&&\int\limits_0^{M_D/m} dy\vec f^{N/D}(y)
= \langle P_D \, S_D \mid
\bar N(0) \gamma_5\gamma_3 N(0)
 \mid P_D \, S_D \rangle = 0.9215.
\label{sr3a}
\end{eqnarray}
This is opposite to the case of the nucleon, where
$h_1^N$ is expected~\cite{jj} to be larger than $g_1^N$
(after exclusion of the quark charge factors).
The effect is different, since in the polarized
deuteron both nucleons are essentially polarized
along the same direction, while in the polarized nucleon
one quark is polarized in opposite direction and
somehow cancels contribution of one of the quarks polarized along
the nucleon polarization.
The difference between $H_1^{N/D}(y)$  and $\vec f^{N/D}(y)$,
 and between the
tensor and axial charges, (\ref{sr2a}) and (\ref{sr3a}), is extremely small
since the deuteron in essentially a nonrelativistic
system and matrix elements of the operators
$\gamma_5\gamma_3$ and $\gamma_5\gamma_3\gamma_0$ are
{\em exactly} the same in the static limit.

The numerical value in eq.~(\ref{sr2a}) is the deuteron
structure factor renormalizing the nucleon tensor charge in the deuteron,
calculated in a particular model of the $NN$-interaction~\cite{tjond}.
Taking into account the differences among existing models, we can
estimate a possible ``model'' error in~(\ref{sr2a}):
$\langle P_D \, S_D \mid
\bar N(0) \gamma_5\gamma_3\gamma_0 N(0)
 \mid P_D \, S_D \rangle = 0.93\pm 0.015$.

To calculate the realistic structure function $h_1^D(x)$ we need
 the nucleon structure functions $h_1^N(x)$. However, so far there is
no existing experimental data for this function, and very little is known
about the form of $h_1^N$ in theory.
In the present paper
we follow the ideas  of ref.~\cite{jj} to estimate $h_1^N$.
Since the sea quarks do not contribute to $h_1^N$, its
flavor content is
simple:
\begin{eqnarray}
h_1^N (x)  = \delta u(x) +\delta d(x),
\label{hn1}
\end{eqnarray}
where $\delta u(x) $ and $\delta d(x)$ are the contributions
 of the u-
and d-quarks, respectively~\cite{jj,hj}.
Since the matrix elements of the operators $\propto \gamma_5 \gamma_3$
and
$\propto \gamma_5 \gamma_3\gamma_0$ coincide in the static limit,
 as a crude estimate we can expect that
\begin{eqnarray}
\delta u(x) \sim \Delta u(x), \quad \quad
\delta d(x) \sim \Delta d(x), \label{dud1}
\end{eqnarray}
where $\Delta u(x)$ and $\Delta d(x)$ are contributions
of the u- and d-quarks to the spin of the nucleon, which is
measured through the
structure function $g_1^N$. Subsequently, the simplest estimate
for $h_1^N$
\begin{eqnarray}
h_1^N (x)  = \alpha \Delta u(x) +\beta \Delta d(x),
\label{hn2}\\
{\rm with \quad\quad}\alpha = \beta=1 \label{ab1}
\end{eqnarray}
should not be too unrealistic. In fact,
the bag model calculation shows that difference between $\delta q$ and
$\Delta q$ is typically only a few percent~\cite{jj}. This analysis is
mostly a qualitative one, since it is
limited by the case with one quark flavor and does not pretend to describe
phenomenology.

To evaluate possible deviations from the simplistic  choice
of $h_1^N$, (\ref{hn2}) with (\ref{ab1}), we
suggest:
\begin{eqnarray}
\alpha = \delta u/\Delta u, \quad \beta= \delta d/\Delta d, \label{ab2}
\end{eqnarray}
where  $\delta q$  and $ \Delta q$  are the first moments of
$\delta q(x)$  and $ \Delta q(x)$, respectively ($q = u, d$).
For
$\delta u$  and $ \delta d$ we can adopt the results from the QCD sum rules
and the bag model calculations~\cite{hj,hj1}. As to
$\Delta u$  and $ \Delta d$, we
can use  the experimental data analysis~\cite{ellis}
or theoretical results, e.g.
the QCD sum rules results~\cite{hhk}.
Thus, we estimate
\begin{eqnarray}
\alpha = 1.5 \pm 0.5, \quad \beta= 0.5 \pm 0.5, \label{ab3}
\end{eqnarray}
at the scale of $Q^2=1$~GeV$^2$.

The realistic form of the distributions $\Delta u(x)$ and $\Delta d(x)$
can be taken from a fit to the experimental data for $g_1^N$. In our
calculations we use parametrization from ref.~\cite{shaf}.
At this point we have to realize that, in spite of the expected
relations~(\ref{dud1}), distributions $\delta q$ and
$\Delta q$ are very different in their nature. Especially at
$x {\ \lower-1.2pt\vbox{\hbox{\rlap{$<$}\lower5pt\vbox{\hbox{$\sim$}}}}\ }0.1$,
where $\Delta q$ probably contains a singular contribution
of the polarized sea quarks, but $\delta q$ does not.
Therefore we expect eq.~(\ref{hn2}) to be a reasonable
estimate in the region of the valence quarks  dominance, say
$x {\ \lower-1.2pt\vbox{\hbox{\rlap{$>$}\lower5pt\vbox{\hbox{$\sim$}}}}\ }0.1$
(see also~\cite{it}).
 For  completely consistent analysis, the parameters $\alpha$ and $\beta$, and
 the distributions $\Delta u(x)$ and $\Delta d(x)$
should be scaled to the same value of $Q^2$. However,
for the sake of the unsophisticated estimates we do not
go into such details.

The results of calculation of the nucleon  and deuteron
 structure functions, $h_1^N$ (solid lines) and $h_1^D$ (dashed lines),
are shown in  Fig.~\ref{h-1d}.
The group of curves 1 represents case (\ref{ab1}), which
is a possible lower limit   for   $h_1^{N,D}$
in accordance with
our estimates (\ref{ab3}). Curves 2 represent the case
  $\alpha = 1.5,\quad\beta = 0.5$, which is close
to the mid point results of the bag model and the QCD sum rules.
The upper limit corresponding to the estimates (\ref{ab3})
is presented by curves 3. For all cases
the deuteron
 structure function is suppressed compared to the one for the nucleon, mainly
due to the depolarization effect of the D-wave in the deuteron.
This is quite similar to the case of the structure functions $g_1^N$
and $g_1^D$. To illustrate this similarity we show the ratio
of the structure functions $h_1^D/h_1^N$ (solid line)
in Fig.~\ref{rat} together with the ratio $g_1^D/g_1^N$ (dotted line).
The straight dash-dotted line in  Fig.~\ref{rat} represents
the value of the matrix elements (\ref{sr2a}) and (\ref{sr3a}), which
are not
distinguishable on this scale. This line approximately
corresponds to
eqs.~(\ref{sr4}) and (\ref{sr5}).

Note that our  estimate of the nucleon structure function $h_1^N$,
(\ref{ab3}),
gives systematically a larger value for the function than naive suggestion
(\ref{ab1}),
the curves 1 in Fig.~\ref{h-1d},
which essentially corresponds to the estimate $h_1^N \simeq (18/5) g_1^N$,
neglecting possible negative contribution of the s-quark sea~\cite{ellis}.
The large size of the effect suggests that it can be detected in
future experiments
with the deuterons~\cite{exp}.
The ratio
of the structure functions $h_1^D/h_1^N$ in this case is also
shown
in Fig.~\ref{rat} (dashed line).

\section{Summary}

We have considered  the chiral-odd structure function of  twist
two,
$h_1^D(x)$, and its first moment,
the tensor charge of the deuteron.
In particular,
\begin{enumerate}
\item The structure function of the deuteron is calculated
in the Bethe-Salpeter formalism for the deuteron amplitude.
Explicit analytical and numerical results for the
effective distribution function of the nucleons in the deuteron,
defining the deuteron structure function $h_1^D$ are obtained.
\item It is shown that the deuteron structure function $h_1^D(x)$
exhibits deviation from the nucleon structure function $h_1^N$ similar
to the case of the usual spin-dependent
structure functions $g_1^D$ and $g_1^N$. The deuteron
structure factor renormalizing the tensor charge of the nucleon
in the deuteron is calculated.
\item A simple phenomenologically motivated model for the
evaluation of the nucleon
function $h_1^N$ is suggested, using the available results
of the QCD sum rules and the bag model for $h_1^N$. The estimates
show  that the nucleon structure function, $h_1^N(x)$,
can be significantly larger than the spin-dependent
structure function after extraction of the averaged
quark charge, i.e.
$ (18/5) g_1^N(x)$.
This effect definitely can be tested experimentally in the semi-inclusive deep
inelstic scattering
and, therefore,
can bring a better understanding of the spin structure of the nucleon.
\end{enumerate}

\section{Acknowledgements}

H.X. He would like to thank X. Ji for useful discussion during his
visit at MIT. This work in part is supported by NSERC, Canada, and
INFN, Italy.
H.X. He is also supported in part by the National Natural
Science Foundation of China  and Nuclear Science Foundation of China.

\newpage

\centerline{\large \bf Figure captions}

\vskip 1cm

 \begin{figure}[h]
\caption{The effective distribution function $H^{N/D}(y)$ (solid line)
for the nucleon
contribution to the deuteron structure function $h_1^D(x)$.
Other effective distributions: spin-dependent
 $\vec f^{N/D}(y)$ (dashed line)
and spin-independent $f^{N/D}(y)$ (dotted line).
}
\label{H}
\end{figure}

\begin{figure}[h]
\caption{The structure functions $h_1^N(x)$ (solid lines) and $h_1^D(x)$.
Groups of curves:
1 - $\alpha = 1.0,\beta = 1.0$ (``lower limit'');
2 - $\alpha = 1.5,\beta = 0.5$ (``mid point'');
3 - $\alpha = 2.0,\beta = 0.0$(``upper limit'') (see eqs.~(\ref{hn2}) and
(\ref{ab3})).
}
\label{h-1d}
\end{figure}

\begin{figure}[h]
\caption{The ratio of the deuteron and nucleon structure functions
$h_1^D(x)/h_1^N(x)$. Curves: solid line - $\alpha = 1.5,\beta = 0.5$;
dashed line - $\alpha = 1.0,\beta = 1.0$. The dotted curve presents the
ratio of the spin-dependent structure functions $g_1^D(x)/g_1^N(x)$.
The dash-dotted line approximately
corresponds to
eqs.~(\ref{sr4}) and (\ref{sr5}).}
\label{rat}
\end{figure}

\phantom{.}

\newpage

\let\picnaturalsize=N
\def\picsize{15.cm}
\def\picfilename{fig1.eps}
\ifx\nopictures Y\else{\ifx\epsfloaded Y\else\input epsf \fi
\let\epsfloaded=Y
\centerline{\ifx\picnaturalsize N\epsfxsize \picsize\fi
\epsfbox{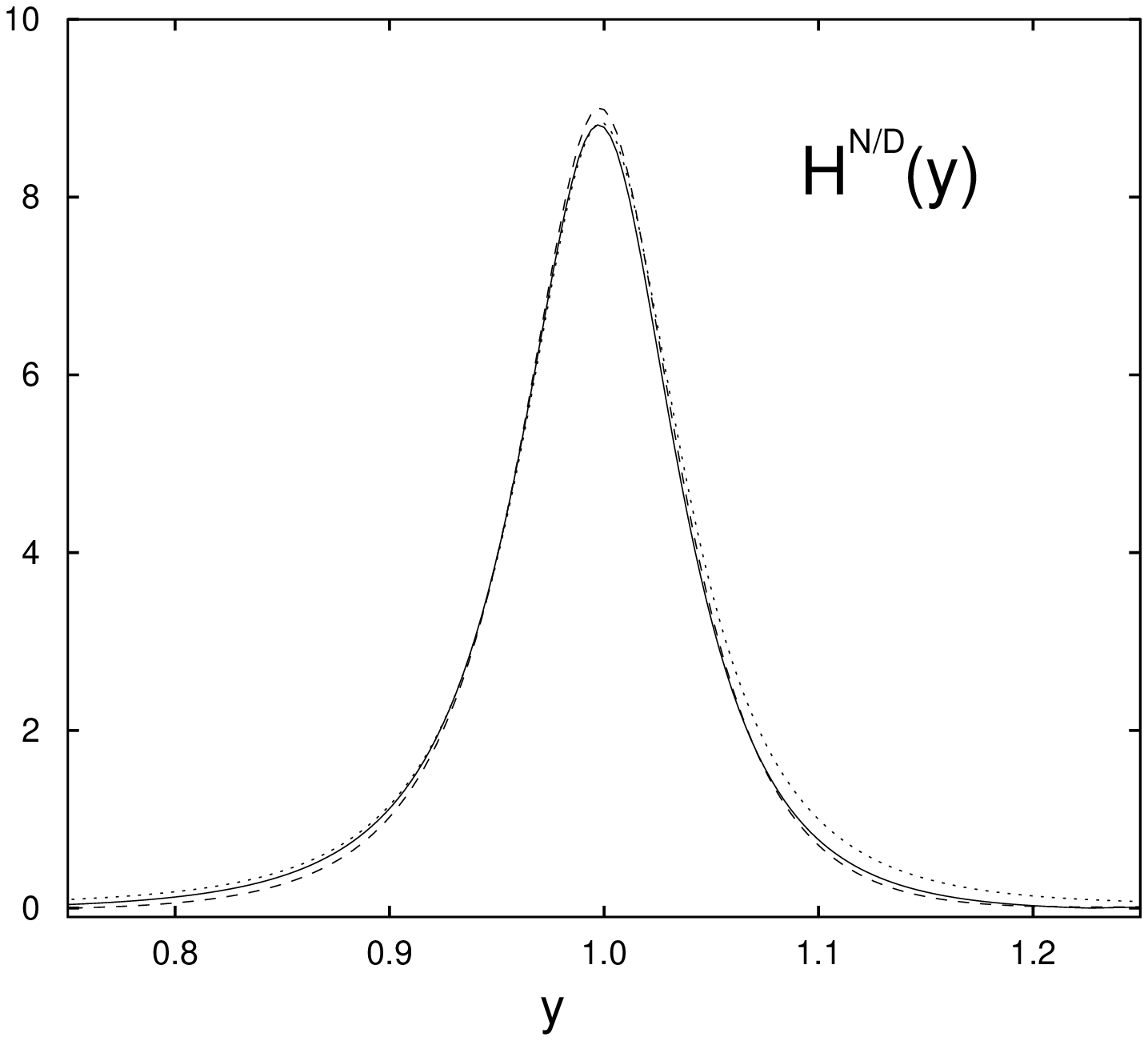}}}\fi

\vfill
Fig.~\ref{H}.  A. Umnikov et al, The chiral-odd structure function...

\newpage

 \let\picnaturalsize=N
\def\picsize{12cm}
\def\picfilename{fig2.eps}
\ifx\nopictures Y\else{\ifx\epsfloaded Y\else\input epsf \fi
\let\epsfloaded=Y
\centerline{\ifx\picnaturalsize N\epsfxsize
 \picsize\fi \epsfbox{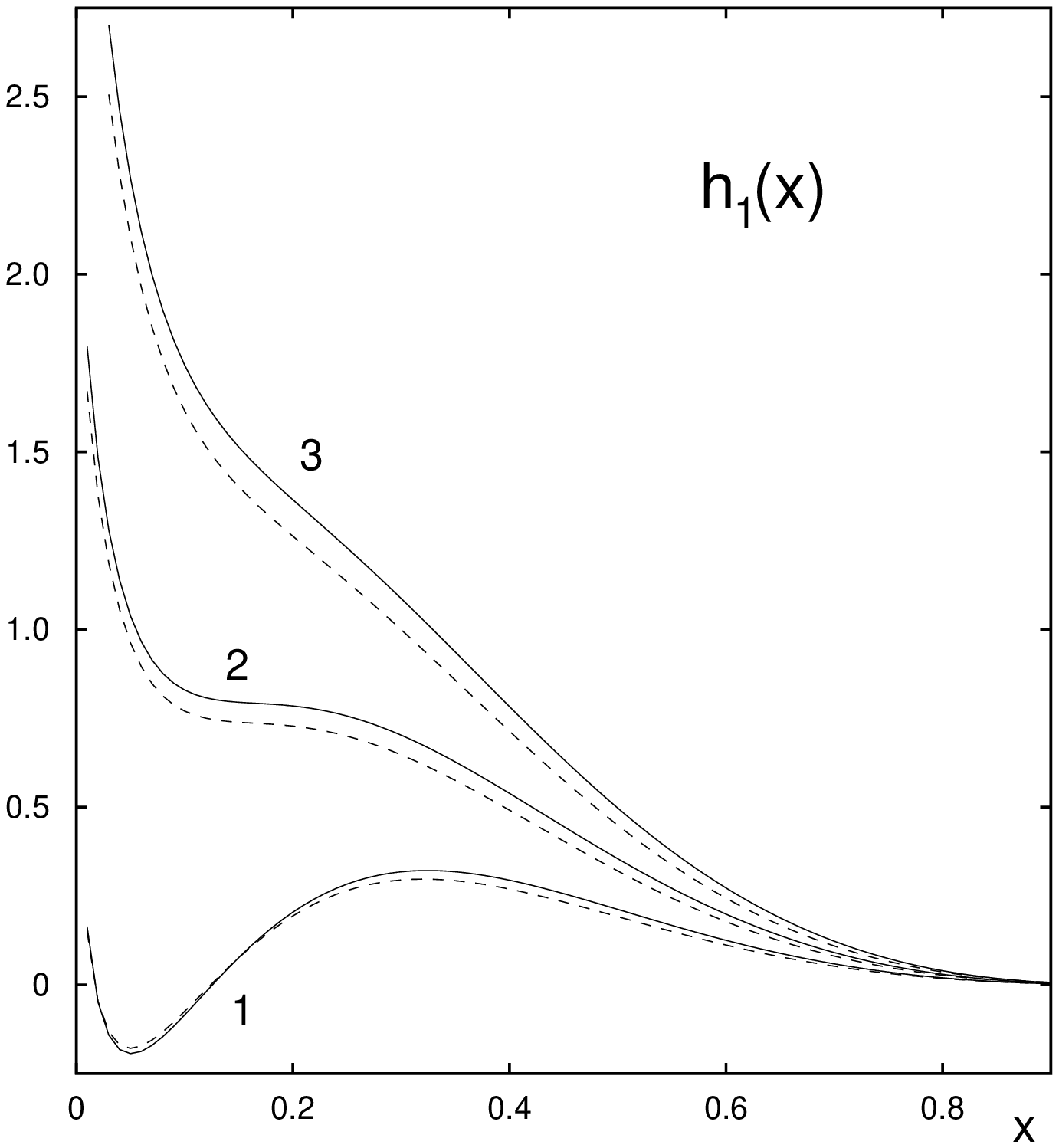}}}\fi
\vfill
Fig.~\ref{h-1d}.  A. Umnikov et al, The chiral-odd structure function...

\newpage

 \let\picnaturalsize=N
\def\picsize{12cm}
\def\picfilename{fig3.eps}
\ifx\nopictures Y\else{\ifx\epsfloaded Y\else\input epsf \fi
\let\epsfloaded=Y
\centerline{\ifx\picnaturalsize N\epsfxsize
 \picsize\fi \epsfbox{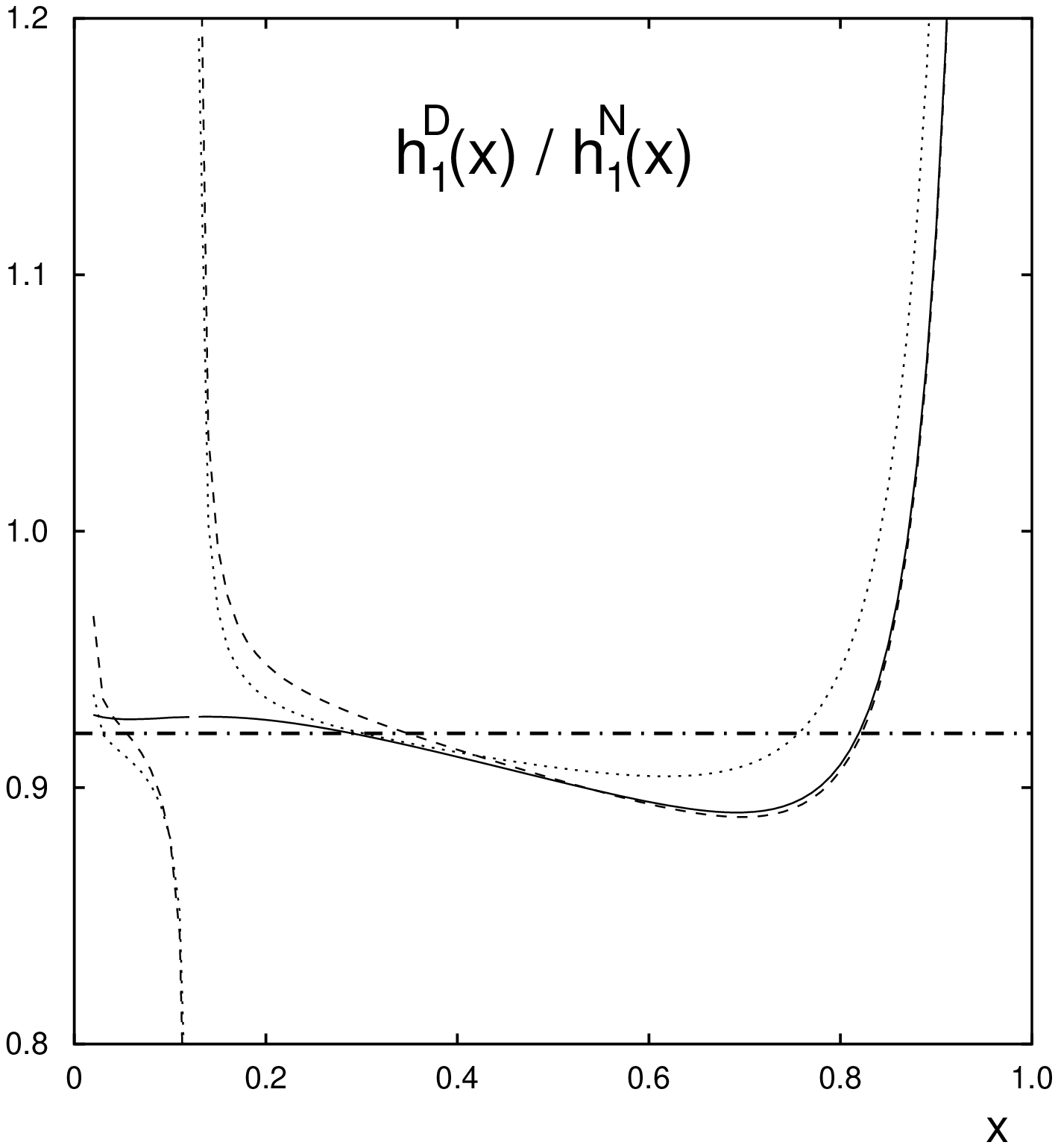}}}\fi
\vfill
Fig.~\ref{rat}.  A. Umnikov et al, The chiral-odd structure function...

 \end{document}